\documentclass{aastex6}
\usepackage{mathtools}
\usepackage{amsmath}
\usepackage{float}
\usepackage{longtable}
\usepackage{soul}

\DeclareMathOperator{\sinc}{sinc}
\fullcollaborationName{The Friends of AASTeX Collaboration}

\begin{document}


\title{Establishing $\alpha$ Oph as a Prototype Rotator: Precision Orbit with new Keck, CHARA, and RV Observations}
%% Use \author, \affil, plus the \and command to format author and affiliation 
%% information.  If done correctly the peer review system will be able to
%% automatically put the author and affiliation information from the manuscript
%% and save the corresponding author the trouble of entering it by hand.
%%
%% The \affil should be used to document primary affiliations and the
%% \altaffil should be used for secondary affiliations, titles, or email.

%% Authors with the same affiliation can be grouped in a single

\author{Tyler Gardner\altaffilmark{1}, John D. Monnier\altaffilmark{1}, Francis C. Fekel\altaffilmark{2}, Michael Williamson\altaffilmark{2}, Fabien Baron\altaffilmark{3}, Sasha Hinkley\altaffilmark{4}, Michael Ireland\altaffilmark{5}, Adam L. Kraus\altaffilmark{6}, Stefan Kraus\altaffilmark{4}, Rachael M. Roettenbacher\altaffilmark{7}, Gail Schaefer\altaffilmark{3}, Judit Sturmann\altaffilmark{3}, Laszlo Sturmann\altaffilmark{3}, Theo Ten Brummelaar\altaffilmark{3}}

\altaffiltext{1}{Astronomy Department, University of Michigan, Ann Arbor, MI 48109, USA}
\altaffiltext{2}{Center of Excellence in Information Systems, Tennessee State University, Nashville, TN 37209, USA}
\altaffiltext{3}{The CHARA Array of Georgia State University, Mount Wilson Observatory, Mount Wilson, CA 91203, USA}
\altaffiltext{4}{University of Exeter, School of Physics and Astronomy, Stocker Road, Exeter, EX4 4QL, UK}
\altaffiltext{5}{Research School of Astronomy \& Astrophysics, Australian National University, Canberra ACT 2611, Australia}
\altaffiltext{6}{Department of Astronomy, University of Texas at Austin, Austin, TX, 78712, USA}
\altaffiltext{7}{Yale Center for Astronomy and Astrophysics, Department of Physics, Yale University, New Haven, CT 06520, USA}

\begin{abstract}
Alpha Ophiuchi (Rasalhague) is a nearby rapidly rotating A5IV star which has been imaged by infrared interferometry. $\alpha$ Oph is also part of a known binary system, with a companion semi-major axis of $\sim$430 milli-arcseconds and high eccentricity of 0.92. The binary companion provides the unique opportunity to measure the dynamical mass to compare with the results of rapid rotator evolution models. The lack of data near periastron passage limited the precision of mass measurements in previous work. We add new interferometric data from the MIRC combiner at the CHARA Array as well as new Keck adaptive optics imaging data with NIRC2, including epochs taken near periastron passage. We also obtained new radial velocities of both components at Fairborn Observatory. Our updated combined orbit for the system drastically reduces the errors of the orbital elements, and allows for precise measurement of the primary star mass at the few percent level. Our resulting primary star mass of $2.20\pm0.06$ M$_{\odot}$ agrees well with predictions from imaging results, and matches evolution models with rotation when plotting on an HR diagram. However, to truly distinguish between non-rotating and rotating evolution models for this system we need $\sim$1\% errors on mass, which might be achieved once the distance is known to higher precision in future Gaia releases. We find that the secondary mass of $0.824\pm0.023$ M$_{\odot}$ is slightly under-luminous when compared to stellar evolution models. We show that $\alpha$ Oph is a useful reference source for programs that need $\pm$1 milli-arcsecond astrometry.
\end{abstract}




\section{Introduction} 
\label{sec:intro}
Observational methods that provide fundamental properties of stars are crucial for benchmarking stellar evolution models. Binary stars are frequently targeted since they provide a direct measurement of stellar mass to compare with models. Visual binary orbits alone provide the sum of masses in the system through Kepler's laws, and combining visual orbits with velocity information from double-lined spectroscopic orbits gives orbital parallax and masses of individual components of the system. Imaging of stellar surfaces is another important measurement which can reveal oblateness, latitude dependencies on radius / temperature, and spots and other surface features (for a review on imaging of stars see e.g. \citealt{vanbelle2012}). Due to large distances, however, most stars are unresolved point sources to traditional single aperture telescopes. Long-baseline optical interferometers with $<$1 milli-arcsecond (mas) resolution are needed to image the largest stars in the sky. Rapidly rotating stars are particularly interesting targets for imaging with optical interferometers, since imaging can provide measurements of inclination, gravity darkening, differential rotation, as well as stellar mass and age estimates (e.g. \citealt{monnier2007,monnier2012,zhao2009,desouza2014}). 

$\alpha$ Oph (Rasalhague, HD 159561) is a nearby, bright, A5IV star which is both a rapid rotator (rotating at $\sim$90\% of its breakup velocity) and in a $<$1 arcsecond visual binary system. This combination provides a unique opportunity to benchmark mass estimates from imaging models of rapid rotators to the direct measurement provided from the binary orbit. \citet{mccarthy1983} was the first to resolve the $\sim$8.6-year secondary component of $\alpha$ Oph with speckle interferometry, and the orbit has been monitored since then. \citet{hinkley2011} carried out a thorough investigation of the visual orbit available at the time from adaptive optics (AO) imaging, combined with photometry and photocenter motion in order to obtain a measurement of both components' masses at $\sim$10\% uncertainty level. Their masses for the A and B components of 2.40$^{+0.23}_{-0.37}$ M$_{\odot}$ and 0.85$^{+0.06}_{-0.04}$ M$_{\odot}$ were lower than previous measurements \citep{gatewood2005}. These lower mass values were in better agreement with the results provided by rapid rotator models from interferometric imaging with the Michigan Infra-Red Combiner (MIRC) at the Georgia State University Center for High Angular Resolution Astronomy (CHARA) Array, which estimated a mass of 2.1$\pm$0.02 M$_{\odot}$ for the primary using stellar evolution models, or 1.7-2.2 M$_{\odot}$ using their new oblateness model method \citep{zhao2009}. \citet{bailey2020} utilized the polarization signal arising from rapid rotation in order to estimate a mass of 2.0$\pm$0.4 M$_{\odot}$ for the primary, also consistent with the lower mass of \citet{hinkley2011}. $\alpha$ Oph has also been targeted for asteroseismic modeling efforts \citep{monnier2010}, in order to learn more about the interior of stars rotating near breakup velocity. Since mass is critical to understanding a star's evolution and physical properties, it is imperative that we precisely measure the mass of well-studied systems when possible.

Though the binary orbit semi-major axis is $\sim$400 milli-arcseconds (mas), the separation of the components is $<$20 mas near periastron since eccentricity is very high at $e=0.92$. Up until this work there was no orbital data near periastron passage, which is crucial to measure mass at the few percent precision level needed to check rotator models at high precision. The updated orbit of \citet{hinkley2011} predicted the time of periastron passage, and we obtained new interferometric and AO imaging data near the passage to improve mass precision. We also obtained double-lined radial velocity (RV) data which allows us to directly measure orbital parallax of the system as well as individual masses. Hence our mass measurements are purely dynamical and do not rely on any outside measurement or model-based result. Our new orbit for $\alpha$ Oph serves as a benchmark test of rapidly rotating stellar models. $\alpha$ Oph is bright ($V=2.1$) and easily observable to most telescopes which can observe bright stars. Our new well-covered orbit allows for precise astrometry prediction across all position angles of the orbit, making this target a good standard astrometric source. 

In Section \ref{sec:observations} we describe our new observations from interferometry, AO imaging, and spectroscopy. Section \ref{sec:orbitfitting} details our combined visual + RV orbit fitting model for the binary system. In Section \ref{sec:orbitresults} we present our new visual and spectroscopic orbit along with physical parameters and masses for each component. We plot these stars on an HR diagram in Section \ref{sec:evolution}. Finally, we comment on the orbital precision and use of $\alpha$ Oph as an astrometric reference in Section \ref{sec:astrometric}.

\section{Observations and Data Reduction} 
\label{sec:observations}
\subsection{MIRC at the CHARA Array}
We obtained two previously unpublished epochs of $\alpha$ Oph near periastron in 2012 using the high angular resolution of MIRC. We attempted observations again during the periastron passage of 2020, though poor weather during our observing run prevented us from obtaining any data. MIRC is a H-band combiner of six 1-meter telescopes at the CHARA Array. The CHARA Array is an optical/near-IR interferometer with baselines up to 330 meters \citep{Brummelaar2005}. The MIRC instrument is described in detail in \citet{Monnier2006} and \citet{che2010, che2012}. The MIRC combiner measures visibilities and closure phase of our targets. Calibrator stars are observed between science observations to measure visibility loss due to time-variable factors such as atmospheric coherence time, differential dispersion, and birefringence in the beam train. The MIRC datasets were reduced with the standard MIRC data pipeline in IDL described in previous MIRC papers (e.g. \citealt{monnier2012}).  The star $\gamma$~Oph ($\Theta_{\rm UD} = 0.571\pm0.040$ milliarcseconds; Source: Jean-Marie Mariotti Center Searchcal tool, \citealt{chelli2016}) was used to calibrate the instrumental transfer function for all MIRC $\alpha$~Oph observations presented here. 

For each MIRC night we fit to the following binary star model of complex visibility, $V$, in order to measure binary position \citep[e.g.]{herbison-evans1971,boden1999}:
\begin{equation}
    V = \frac{V_1 + \Gamma f V_2 e^{-2\pi i (u\alpha+v\delta)}}{1+f}.
\label{vis_eqn}
\end{equation}

For determining simply the binary parameters, a detailed image of the nearly ``edge-on" rapidly-rotating rotating primary star \citep[see][]{zhao2009} is not required.  Here, $\alpha$~Oph A was approximated as a uniformly-bright elliptical disk (position angle of -53.88 degrees from \citealt{zhao2009}) and is represented by $V_1$ in Equation~\ref{vis_eqn}.  $\alpha$~Oph B was treated as an unresolved star represented by $V_2$.  Other free parameters in fit were the binary separation in right ascension (R.A., $\alpha$) and declination (DEC, $\delta$); a monochromatic flux ratio between the two components $f$; as well as a bandwidth smearing correction $\Gamma = \sinc[b(u\alpha+v\delta)]$, where $b = 1/R$ and $R$ is the spectral resolution of the spectrograph. As is standard, the location of each datum on the uv-plane is denoted by parameters $u$ and $v$. We fit to the calibrated squared-visibility and the closure phase. We investigated whether it was acceptable to use this simple symmetric intensity model for the primary rather than the full gravity-darkenened oblate spheroid model when measuring the binary separation. We calculated the expected photocenter shift due to the slight asymmetry in the intensity distribution from the gravity darkening effect using the full model of \citet{zhao2009}, and found the model shift is only 4 micro-arcseconds away from the true center of mass, a small systematic error that we can safely ignore here. Errors on astrometry are estimated by deriving a $\chi^2$ surface for a grid in relative R.A. and DEC and finding the 1-$\sigma$ confidence contour. To remain consistent with the AO imaging data, we convert this confidence contour to an error in position angle and separation. Results from the two MIRC epochs, along with the rest of the astrometry data, are presented in Table \ref{table:astrometry-alpoph}. 

\subsection{Keck Adaptive Optics Imaging}
%We include 6 previously unpublished epochs of $\alpha$ Oph with the Keck-II adaptive optics imager NIRC2 between 2002-2014.   
%For four of the six epochs, the $\alpha$~Oph components are well resolved by the NIRC2 system with the adaptive optics only. In these cases, astrometry was extracted by fitting a 2-dimensional gaussians to the core of the two components. Astrometric errors were determined from scatter in these positions using multiple data frames per epoch.  Using these sub-pixel centroid positions, the geometric distortion of NIRC2 was corrected using the algorithm described in \citet{yelda2010} and then the relative separation and position were reported in Table~\ref{table:astrometry-alpoph}.
The $\alpha$ Oph system was observed with the Keck-II adaptive optics system and the facility AO imager NIRC2 in six previously-unpublished epochs obtained between 2002 March 27 and 2014 June 10. All observations were taken using the narrow-band camera with several different choices of filters, typically using a subarray to shorten exposure times and minimize saturation. We processed these images following the general procedures described in \citet{kraus2016}, performing a linearity correction, subtracting mode-matched dark frames, and dividing by the most contemporaneous flatfield available for the filter used in each observation. For the imaging observations, we also performed ``destriping'' to remove spatially correlated readnoise that is mirrored in the 4 quadrants of the NIRC2 detector, as well as identifying cosmic rays and interpolating over them. Finally, we flagged all saturated pixels so they would not be used in our PSF-fitting analysis.

For four of the six epochs, the $\alpha$~Oph components are well-resolved in images and could be fit with a $\chi^2$ minimization of a double-source model. Following \citet{kraus2016}, for each frame we iteratively fit the separation, PA, and contrast of the binary pair, and then tested the 1000 most contemporaneous single-star images from the archive to identify the optimal empirical template. We repeated these steps until the same PSF template yielded the lowest $\chi^2$ for two consecutive iterations. Using the pixel positions corresponding to these astrometric measurements, the geometric distortion of NIRC2 was corrected using the algorithm described in \citet{yelda2010} and then the relative astrometry was reported in Table~\ref{table:astrometry-alpoph}. For each epoch, we report the mean separation and PA for all frames in the given filter. The uncertainty represents the RMS of the observed frames, added in quadrature with the systematic uncertainty in separation ($\sigma_{\rho} = 1.4$ mas) and in PA ($\sigma_{PA} = \arctan(\frac{1.4 mas}{\rho})$) that results from the residual uncertainty in the distortion solution of \citet{yelda2010}. The systematic uncertainties in the pixel scale and detector orientation are negligible compared to this term.

Two of the archival epochs (both in 2012 April) occurred very close to periastron passage with separation less than the diffraction limit of the telescope.  In order to recover high-precision separations in this situation, the 18-hole aperture mask within NIRC2 was used following procedures discussed in other recent papers \citep[e.g.,][]{ireland2008b,kraus2012,willson2016,rizzuto2020}.  By fitting a precise binary model to the interferometric visibilities and closure phases formed by the aperture mask, a precise component separation can be extracted after fixing the component flux ratio established from previous wide-separation measurements. For 2012 Apr 5 we have a measurement taken both at Keck and at the CHARA Array which show good agreement within errors of the binary position taken with two different instruments.

\subsection{Fairborn Observatory Radial Velocities}
Between 2011 and 2020 we obtained 145 new radial velocity 
(RV) data points for the primary component of $\alpha$ Oph,
and 107 RVs for the secondary. This period covers two periastron
passages of $\alpha$ Oph, crucial for obtaining masses at the
few percent level. These data were taken with the Tennessee
State University 2~m Automated Spectroscopic Telescope (AST)
and its echelle spectrograph at the Fairborn Observatory in 
southeast Arizona \citep{ew07}. The detector was a
Fairchild 486 CCD that has a 4K $\times$ 4K array of 15 
$\mu$m pixels \citep{fetal13}. The spectra have a resolution
of 0.24~\AA, corresponding to a resolving power of 25000 at
6000~\AA, and cover a wavelength range from 3800 to 8260~\AA.
After acquiring a couple test spectra of $\alpha$ Oph and
trying to measure RVs, we settled on the following observing
sequence to optimize RV measurement of the very broad lined
primary star and the extremely weak lines of its very faint
companion. Because of the brightness of the $\alpha$~Oph system,
we typically acquired 40 consecutive 30 sec observations,
which we then summed together into a single spectrum that 
has a significantly improved signal-to-noise ratio (SNR).
 
\begin{figure}[H]
\centering
\includegraphics[width=4.5in]{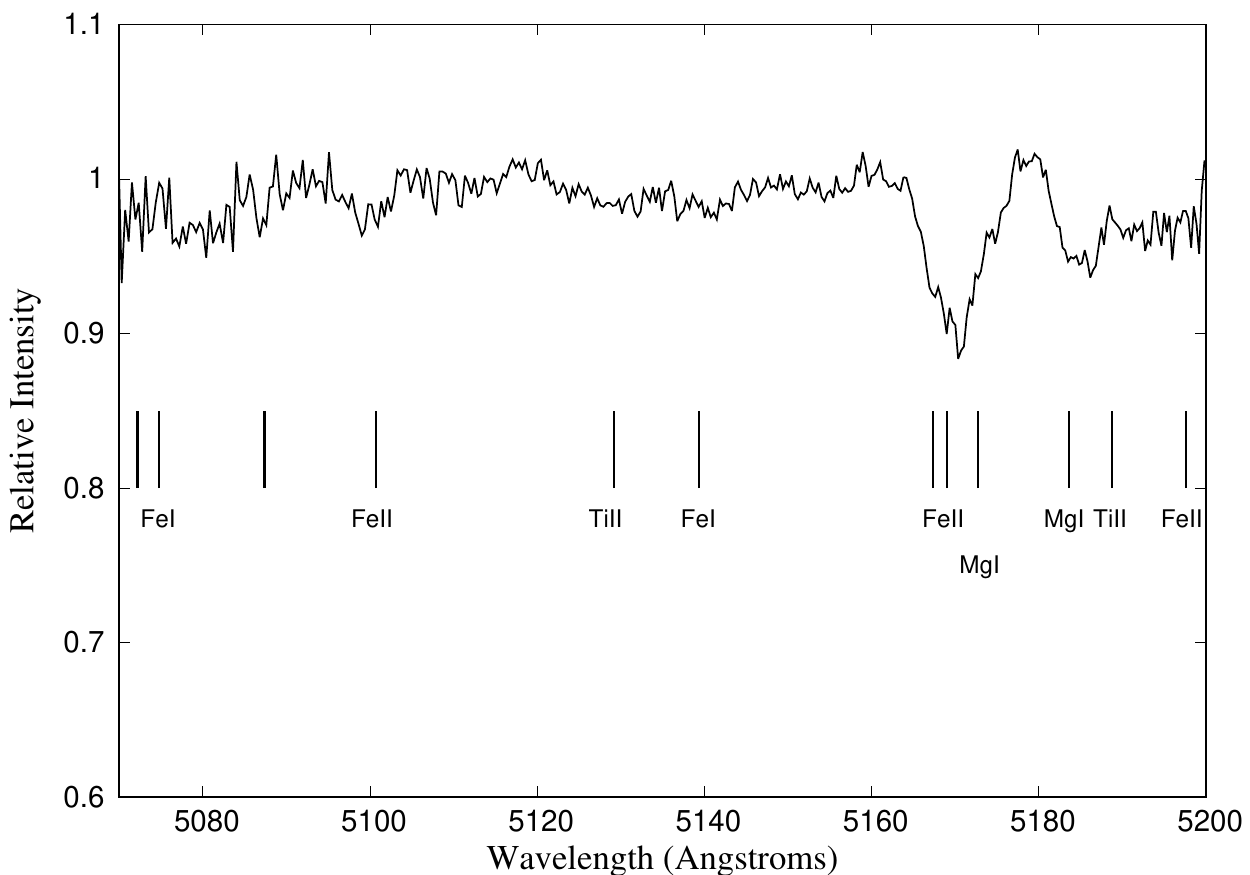}
\caption{We show an echelle order centered at about 5135~\AA, for an $\alpha$ 
Oph spectrum taken with the 2m AST at Fairborn Observatory. Due to the 
high rotational velocity of the A star and blending issues with nearby 
lines, we were unable to use our A-star lines list to obtain RV. Tick 
marks indicate the rest positions of individual lines in the list. 
Though the Mg I lines appear relatively strong in this order, they are 
unusable due to blending problems with nearby lines. The other lines 
are too weak to measure reliably. As discussed in the text, we instead 
determined RVs from the H$\beta$ line of the primary. }
\label{alpoph_spectrum}
\end{figure}

For the A5~IV primary we initially tried to measure the
velocities of the lines compiled in our A star line list.
This list primarily consists of singly ionized elements
of Fe and Ti. Unfortunately, we were unable to obtain
usable RVs with the lines in this list due to a combination
of factors. The most problematic one is the very high rotational
velocity of the A star, 228 km~s$^{-1}$ \citep{royer2002},
which causes the lines to be extremely broad and shallow.
This makes it much more difficult to measure the velocity
centers of the profiles compared with the measurement of a
narrow-lined star. Another significant problem is that the
very large rotational broadening of the lines greatly increases 
the likelihood of blending with nearby lines. These problems 
are illustrated in Figure \ref{alpoph_spectrum}. Shown is an 
echelle order centered at about 5135~\AA. Tick marks indicate 
the rest positions of individual lines in our A star line list. 
While this order includes the relatively strong Mg I lines, 
they are unusable because of blending problems and the 
other lines are extremely weak. Additional problems with RV 
measurement, exacerbated by the large rotational broadening, 
occur because of difficulties with continuum rectification 
and the fact that the ends of the echelle orders have lower SNRs. 

Thus, instead of velocities determined from lines in the A
star list, we individually measured the hydrogen lines, H$\alpha$, 
H$\beta$, and H$\gamma$. In addition to producing higher SNR, 
summing the consecutive spectra smoothed out the rapidly varying 
ripples in the hydrogen profiles that are presumably caused by
 pulsation. Of the three lines, H$\gamma$ is well into the 
 blue part of the spectrum where the throughput of our
 echelle system and detector is low, and there are an
 increasing number of metal lines that cause blending
 problems. H$\alpha$ is near the end of two echelle orders,
 where there is a lower SNR, and there are increased problems
 with continuum rectification. The H$\beta$ line is closer 
 to the middle of its echelle order than H$\alpha$ is, and
 RVs of H$\beta$ result in the orbit that is most consistent
 with the secondary velocities. Thus, for the primary we
 choose to adopt the RVs determined from the H$\beta$ line.
 However, we note that continuum rectification problems can
 also affect this broad line and may be at least partly
 responsible for the significant center-of-mass velocity
 difference that we find between its orbit and that of the
 secondary.
  
 \citet{ftw09} provided a general explanation of our usual
 velocity measurement procedure. In particular, for the 
 secondary of $\alpha$ Oph we used our solar-type star line
 list with which we were able to detect the extremely weak
 secondary component. That component has a mean line depth 
 of just 0.002 and so is barely detectable in most of our
 summed spectra. The solar list contains 168 lines in the
 wavelength range 4920--7100~\AA. Given the weakness of
 average line profile, we chose to report RVs as whole
 numbers. RVs for the primary and secondary components are
 reported in Table \ref{table:rv-alpoph}.

%%%%%%%%%%%%%%%%%%

\section{Orbit Fitting}
\label{sec:orbitfitting}
Once we have our measured binary separations and position angles for each night, we are able to fit a Keplerian orbit to the data. The Campbell elements ($\omega$, $\Omega$, $e$, $i$, $a$, $T$, $P$) describe the Keplerian motion of one star of a binary system relative to the other. Those symbols have their usual meanings \citep[e.g.]{Wright2009} where $\omega$ is the longitude of the periastron, $\Omega$ is the position angle of the ascending node, $e$ is the eccentricity, $i$ is the orbital inclination, $a$ is angular semi-major axis, $T$ is a time of periastron passage, and $P$ is orbital period. When including RV data, we also fit to the two semi-amplitudes $K_A$ and $K_B$ and system velocity $\gamma$. Note that as described in Section \ref{sec:observations}, there are two rather different system velocities for the two components due at least in part to continuum normalization because of the use of the very broad hydrogen spectral line to compute the RVs of the primary. The longitude of periastron $\omega$ is traditionally reported for the secondary when fitting to visual binary orbits alone. The convention when combining RV orbits is to report $\omega$ of the primary, which is flipped by 180$^{\circ}$. For visual orbits, there is a 180 degree ambiguity between $\omega$ and $\Omega$. Our RV information breaks this degeneracy. 

For nonlinear least-squares fitting, we use the Thiele-Innes elements to describe our Keplerian orbits. As described in \citet{Wright2009}, these elements convert ($\omega$, $\Omega$, $i$, $a$) to linear parameters (A, B, F, G). We use the Python $lmfit$ package for non-linear least-squares fitting of our data \citep{newville2016}. The Python $astropy$ \citep{astropy,astropy2018} package is also extensively used in our fitting routines. Error bars for the fitted orbital parameters are normally estimated in $lmfit$ from the covariance matrix, but since the orbital elements $P$ and $e$ are nonlinear we instead determine posterior distributions on our orbital parameters with a Markov chain Monte Carlo (MCMC) fitting routine. We carry out MCMC fitting using the Python package \textit{emcee} developed by \citet{Foreman2013}. We use our best-fit orbital elements as a starting point for our 2*N$_{\rm{params}}$ walkers, where the starting point for each walker is perturbed about its best fit value. We assume uniform priors on all of our orbital elements within physically allowed parameter space (i.e. eccentricity is restricted between 0--1, $\omega$ and $\Omega$ between 0--360$^{\circ}$, inclination between 0--180$^{\circ}$, no negative values allowed for semi-major axis, period, periastron passage, semi-amplitudes). The quoted error bars on our orbital elements in Table \ref{orbitelements:alpoph} are the standard deviations of the posterior distributions, and corner plots of the posteriors for the inner and outer orbits show correlations between parameters. 

\section{New Visual and Spectroscopic Orbit}
\label{sec:orbitresults}

\begin{figure}[H]
\centering
\includegraphics[width=7in]{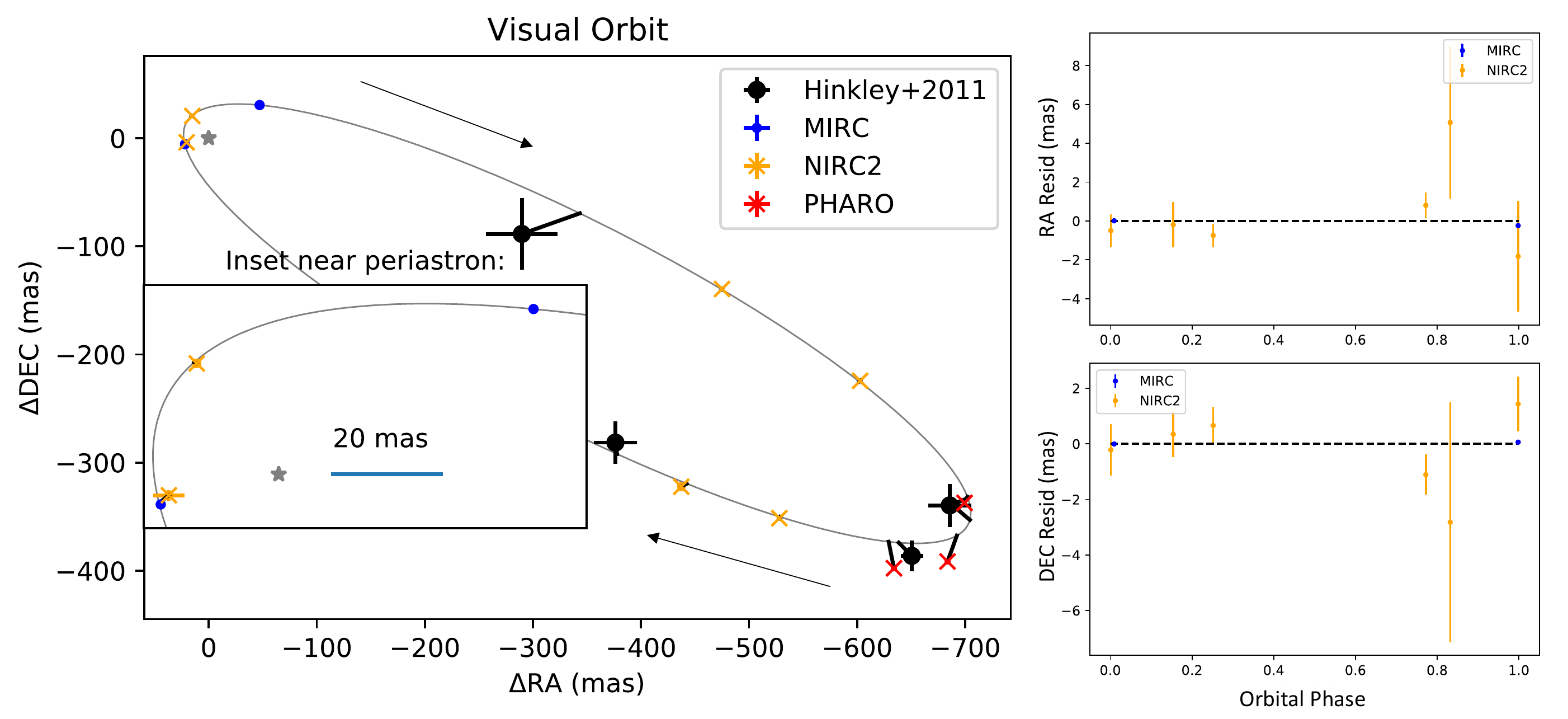}
\caption{We show the visual orbit for the best fit to the combined visual and spectroscopic data of $\alpha$ Oph. The MIRC and NIRC2 epochs near periastron passage, along with new double-lined RV data, allow for a high precision measure of the masses of this system. We also show the \citet{hinkley2011} data for comparison, which were not used in our final orbit fit. All of this data fits our orbit well except for the points taken at Palomar Observatory (particularly, those with the PHARO instrument -- see the text for details). Including this data in the fit has no significant effect on the best-fit orbital elements, which are constrained with our new high precision data. }
\label{alpoph_visual}
\end{figure}

\begin{figure}[H]
\centering
\includegraphics[width=7in]{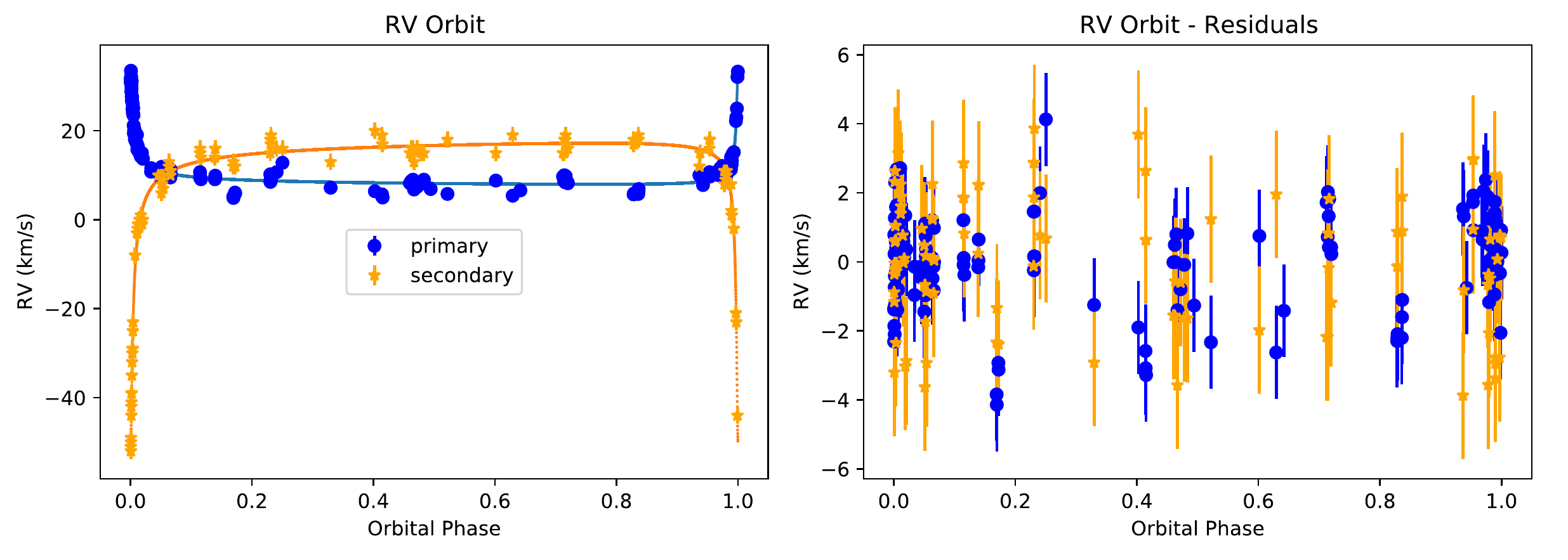}
\caption{We show the RV orbit of the best fit to the combined visual and spectroscopic data of $\alpha$ Oph. As explained in the text, the different lines used to measure RV lead to different velocity offsets. This does not have a significant effect on the orbit, as one can see from the residuals that the offsets are constant across orbital phase. The shape of the RV curve is also heavily set by the combined fit with the visual orbit. }
\label{alpoph_rv}
\end{figure}

Table \ref{table:astrometry-alpoph} compiles all of our previously unpublished data that we use to fit our visual orbit of $\alpha$ Oph. This includes our 2 CHARA/MIRC interferometry and 6 Keck/NIRC2 AO epochs. Table \ref{table:rv-alpoph} presents our new double-line RV data from Fairborn Observatory. In Table \ref{orbitelements:alpoph} we give the best-fit orbital elements when fitting to astrometry alone, RV alone, and the combined fit. We show our best-fit visual orbit in Figure \ref{alpoph_visual} which also plots the data points from \citet{hinkley2011}, and our best fit RV orbit is shown in Figure \ref{alpoph_rv}. For our reported best-fit orbital elements we only fit to our updated high precision data, as this data was observed and analyzed in a systematic and consistent way. Though we checked that most of the historical data points from \citet{hinkley2011} are consistent with our orbit (i.e. within the quoted 1-sigma error values). The exception are the data points in that paper which were taken with Project 1640 and PHARO instruments at Palomar Observatory, which are multiple sigma off from our best-fit orbit. This is not surprising for Project 1640, as the astrometry returned by the instrument was not yet well understood at the time of observation (private communication with authors). The PHARO data points had error values that were much smaller than the rest of the data from that paper, and there is some hint that these errors may be underestimated. \citet{pope2016} measured a new epoch on $\alpha$ Oph from PHARO, and Table 1 of that paper shows that the astrometry solution varies by $>$10 mas depending on the fitting method used. This is significantly larger that the $\sim$1-3 mas error bars given in \citet{hinkley2011} for the PHARO data. Figure \ref{alpoph_visual} designates the PHARO data points separately from the rest of the \citet{hinkley2011} data points, to highlight the fact that they do not fit well like the rest of this data. Though we do not include any of this older data in our fit, we checked that including them in the fit does not significantly change our orbital elements or masses within error bars. 

With a combined RV and visual orbit, we are able to compute the distance and masses of both components of the system. We present these values in Table \ref{physicalelements:alpoph}. Our best-fit distance is 14.80$\pm$0.13 pc, which is consistent within error bars with the Hipparcos parallax of 14.90$\pm$0.24 pc \citep{vanLeeuwen2007}. Our best-fit mass values for the primary and secondary are $2.20\pm0.06$ M$_{\odot}$ and $0.824\pm0.023$ M$_{\odot}$ respectively. These mass errors are both at the 2.7\% level, which is a significant improvement on the $\sim$10\% mass errors presented in \citet{hinkley2011}, though our values are consistent with that work. This precision allows for a more thorough check to rapid rotator evolution models. Our mass on the primary is just barely outside the 1-sigma error bar of the $2.1\pm0.02$ M$_{\odot}$ prediction from evolution models using the MIRC imaging results of \citet{zhao2009}. It is consistent with the prediction in that work of 1.7-2.2 M$_{\odot}$ using their oblateness method to estimate mass. Our mass value for the primary also agrees well with polarization work from \citet{bailey2020}, though the precision is lower for that work at $2.0\pm0.4$ M$_{\odot}$. Our mass values for the primary are significantly lower than previous literature values of $2.84\pm0.19$ M$_{\odot}$ \citep{gatewood2005} and 4.9 M$_{\odot}$ \citep{kamper1989}, which did not agree well with the rapid rotator model results. The addition of data near periastron passage is likely the cause of the discrepancy with previous literature orbit results, and is indeed crucial for constraining eccentric binary orbits. 

 Our best-fit orbital elements of $a$, $i$, $e$, $\omega$, and $\Omega$ agree well within error bars with \citet{hinkley2011}, though we note that due to better coverage our values are much more tightly constrained. \citet{hinkley2011} chose to fix their orbital period to 3148.4 days from \citet{gatewood2005}, which is $\sim$9 days larger than our best-fit value of $3139.72\pm0.28$ when combining the new spectroscopic orbit with the visual orbit. Reported orbital period in literature range from 3109--3165 days \citep{kamper1989,augenson1992,gatewood2005}, so our value is consistent with this spread though much more tightly constrained. Corner plots from the posteriors of our MCMC fitting routine are shown in Figure \ref{alpoph_mcmc}. Our errors are derived from the 1-$\sigma$ standard deviation in these posterior distributions, and the plots show correlation between parameters. Since this is an eccentric orbit, there are correlations between $e$ with parameters $\omega$, $\Omega$, and $T$ which can be seen in our plots. 
 
 Similar to \citet{hinkley2011}, we are able to compute the mutual inclination between the primary star's rotation axis and the plane of the binary orbit from the equation of \citet{fekel1981}. Using the values in \citet{zhao2009} of the imaged rotator model and our updated binary orbit, we find a mutual inclination of either $42.7 \pm 0.6^{\circ}$ or $133.0 \pm 0.6^{\circ}$. Though our orbit is fully characterized with RV data included, the degeneracy in mutual inclination comes from the unknown spin polarity of the primary. In either case, the mutual inclination is significantly non co-planar. Combined with the high eccentricity of the binary orbit, this perhaps suggests early interaction in the star formation phases between the binary orbit and the primary rotation angular momentum. 

%astrometry table
\begin{table}[H]
\centering
\caption{$\alpha$ Oph Astrometry}
\label{table:astrometry-alpoph}
\begin{tabular}{lcccccc}
\hline
\colhead{UT Date} & \colhead{JD} & \colhead{sep (mas)}  & \colhead{P.A. ($^\circ$)} & \colhead{Instrument} \\
\hline
%1999 May 6 & 2451304.5 & 770 &  40  &  243.7  &  3.0 & SHARP II &  1 \\
2002 Mar 27 & 2452360.95 & 545.2 $\pm$ 1.9 & 233.61 $\pm$ 0.17 & KECK-II/NIRC2 \\   
%2002 Jul 18 & 2452473.8 & 470 & 20 & 233.2 & 2.4 & VisIm & 1 \\         
%2004 Jun 12 & 2453168.5 & 303   &  33  & 253.0 &  6.3 & Lyot Project & 1 \\  
%2007 Jun 2 & 2454253.9 & 776.5 & 2.1 & 244.6 & 0.4 & PHARO & 1 \\
%2007 Jun 12 & 2454263.5 & 765.2  & 20.0  &  243.65  & 1.50 & Lyot Project & 1 \\  
%2008 Jun 20 & 2454637.5 & 787.8 & 2.8 & 240.6 & 0.4 & PHARO & 1 \\
%2008 Jul 10 & 2454657.5 & 790 & 20 & 239.5 & 1.4 & Project 1640 & 1 \\
%2009 May 12 & 2454963.8 & 756.6  & 7.5  & 239.3  &  1.2 & PUEO & 1 \\
%2009 Jun 20 & 2455002.5 & 748.5 & 2.7 & 238.3 & 0.4 & PHARO & 1 \\
2010 Apr 26 & 2455313.15 & 635.0 $\pm$ 1.5 & 236.70 $\pm$ 0.13 & KECK-II/NIRC2 \\
2012 Apr 5 & 2456022.952 & 22.54 $\pm$ 0.12 & 104.31 $\pm$ 0.08 & CHARA/MIRC \\
2012 Apr 5 & 2456023.05 & 20.7 $\pm$ 2.9 & 100.8 $\pm$ 2.3  & KECK-II/NIRC2 \\
2012 Apr 14 & 2456032.15 & 25.5 $\pm$ 1.0 & 36.5 $\pm$ 1.7 & KECK-II/NIRC2 \\
2012 May 10 & 2456057.939 & 56.19 $\pm$ 0.05 & 302.95 $\pm$ 0.03 & CHARA/MIRC \\
2013 Aug 7 & 2456511.86 & 495.6 $\pm$ 1.5 & 253.87 $\pm$ 0.16  & KECK-II/NIRC2 \\
2014 Jun 10 & 2456819.06 & 641.6 $\pm$ 1.5 & 249.81 $\pm$ 0.13 & KECK-II/NIRC2 \\
\hline
\end{tabular}
\end{table}
%%%%%%%%%%%%%%%%%%

%RV table
\begin{longtable}{lcc}
\caption{$\alpha$ Oph Radial Velocities\tablenotemark{a}}
\label{table:rv-alpoph}\\
\hline
HJD\tablenotemark{b} & $RV_A$ (km~s$^{-1}$)\tablenotemark{c} & $RV_B$ (km~s$^{-1}$)\tablenotemark{d} \\
\hline
\endhead
2455847.6533 & 7.8 & -- \\
2455927.0572 & 10.2 & -- \\
2455930.0570 & 10.0 & -- \\
2455935.0558 & 11.5 & -- \\
2455947.0524 & 12.1 & -- \\
2455957.0473 & 11.9 & -- \\
2455967.0446 & 10.9 & -- \\
2455976.0409 & 10.8 & -- \\
2455991.0334 & 12.4 & -- \\
... & ... & ... \\
\hline
\end{longtable}
\tablenotetext{a}{Full table in Appendix \ref{appendix:rv}}
\tablenotetext{b}{HJD = Heliocentric Julian Date}
\tablenotetext{c}{Errors on primary are 1.4 km~s$^{-1}$.}
\tablenotetext{d}{Errors on secondary are 2 km~s$^{-1}$.}

%\begin{figure}[H]
%\centering
%\includegraphics[width=7in]{figures/alpoph_orbit.pdf}
%\caption{We fit to the combined visual (left) and spectroscopic (right) data of $\alpha$ Oph. The MIRC and NIRC2 epochs near periastron passage, along with new double-lined RV data, allow for a high precision measure of the masses of this system. We also show the \citet{hinkley2011} data for comparison, which were not used in our final orbit fit. All of this data fits our orbit well except for the points taken at Palomar Observatory (particularly, those with the PHARO instrument -- see the text for details). Including this data in the fit has no significant effect on the best-fit orbital elements, which are constrained with our new high precision data. }
%\label{alpoph_orbit}
%\end{figure}

\begin{table}[H]
\centering
\caption{$\alpha$ Oph: Best fit orbital elements}
\label{orbitelements:alpoph}
\begin{tabular}{lcccc}
\hline
\colhead{} & \colhead{\citet{hinkley2011}} & \colhead{Visual Orbit} & \colhead{RV Orbit} & \colhead{Combined Orbit} \\
\hline
P (d) & 3148.4 (fixed) & $3149.6 \pm 5.6$ & $3139.78 \pm 0.35$ & $3139.72 \pm 0.28$ \\
T (JD) & $2452888 \pm 53$ & $2456028.20 \pm 0.03$ & $2456028.26 \pm 0.27$ & $2456028.220 \pm 0.029$ \\
e & $0.92 \pm 0.03$ & $0.93938 \pm 0.00016$ & $0.9370 \pm 0.0012$ & $0.93912 \pm 0.00013$ \\
$\omega$ (deg) & $162 \pm 14$ & $169.98 \pm 0.24$ & $171.18 \pm 0.66$ & $170.21 \pm 0.23$ \\
$\Omega$ (deg) & $232 \pm 9$ & $236.70 \pm 0.17$ & -- & $236.86 \pm 0.16$ \\
i (deg) & $125^{+6}_{-9}$ & $130.67 \pm 0.07$ & -- & $130.679 \pm 0.067$ \\
a (mas) & $427^{+20}_{-13}$ & $410.59 \pm 0.48$ & -- & $409.8 \pm 0.3$ \\
$K_A$ (km/s) & -- & -- & $12.53 \pm 0.22$ & $12.7 \pm 0.2$ \\
$K_B$ (km/s) & -- & -- & $33.35 \pm 0.44$ & $33.74 \pm 0.35$ \\
$\gamma_A$ (km/s) & -- & -- & $8.86 \pm 0.14$ & $8.91 \pm 0.14$ \\
$\gamma_B$ (km/s) & -- & -- & $14.74 \pm 0.22$ & $14.67 \pm 0.21$ \\
\hline
\end{tabular}
%%\tablenotetext{a}{Here we refer to the number of frames used in the data reduction.  A few recorded frames were unusable due to clouds or poor guiding behind the occulting spot.}
\end{table}

\begin{table}[H]
\centering
\caption{$\alpha$ Oph: Physical Properties}
\label{physicalelements:alpoph}
\begin{tabular}{lcc}
\hline
\hline
distance (pc) & $14.80 \pm 0.13$ & This Work \\
$M_{\rm{A}}$ ($M_{\odot}$) & $2.20 \pm 0.06$ & This Work \\
$M_{\rm{B}}$ ($M_{\odot}$) & $0.824 \pm 0.023$ & This Work \\
$\Delta \rm{mag}$ (Kc) & $3.44 \pm 0.12$ & This Work \\
$\Delta \rm{mag}$ (Jc) & $4.14 \pm 0.08$ & This Work \\
Apparent mag, A (Kband) & $1.684 \pm 0.007$ & This Work+\citet{cohen1999} \\
Apparent mag, B (Kband) & $5.12 \pm 0.12$ & This Work+\citet{cohen1999} \\
Apparent mag, A (Jband) & $1.752 \pm 0.005$ & This Work+\citet{cohen1999} \\
Apparent mag, B (Jband) & $5.89 \pm 0.08$ & This Work+\citet{cohen1999} \\
$T_{\rm{eff, A}}$ (K)\tablenotemark{a} & $8250 \pm 100$ & \citet{zhao2009} \\
$L_A$ (L$_{\odot}$)\tablenotemark{a} & $30.2 \pm 1.3$ & \citet{zhao2009} \\
$R_{\rm{pol,A}}$ & $2.390 \pm 0.014$ & \citet{zhao2009} \\
$R_{\rm{eq,A}}$ & $2.871 \pm 0.020$ & \citet{zhao2009} \\
Age (Gyr) & $0.77 \pm 0.03$ & \citet{zhao2009} \\
\hline
\end{tabular}
\tablenotetext{a}{Averaged over surface}
\end{table}

\section{Comparison with Stellar Evolution Models}
\label{sec:evolution}
One of the major opportunities offered by high precision binary star orbits is a comparison of observations to stellar evolution models. \citet{zhao2009} used interferometric imaging and rotator models to measure inclination, equatorial radius and temperature, and rotation speed of $\alpha$ Oph A. Using the true effective temperature and luminosity from those models, we show the position of component A on an HR diagram to compare with its stellar track. We use rotator models from \citet{ekstrom2012} to compute isochrones and stellar tracks for the primary, using a rotation rate of $v/v_{\rm{crit}}=0.9$ consistent with the high rotation rate in \citet{zhao2009}. We also use Mesa Isochrones and Stellar Tracks (MIST) evolution models to compute tracks without rotation \citep{mist1,mist2,mist3,mist4}. We assume a solar metallicity in both models.

\begin{figure}[H]
\centering
\includegraphics[width=7in]{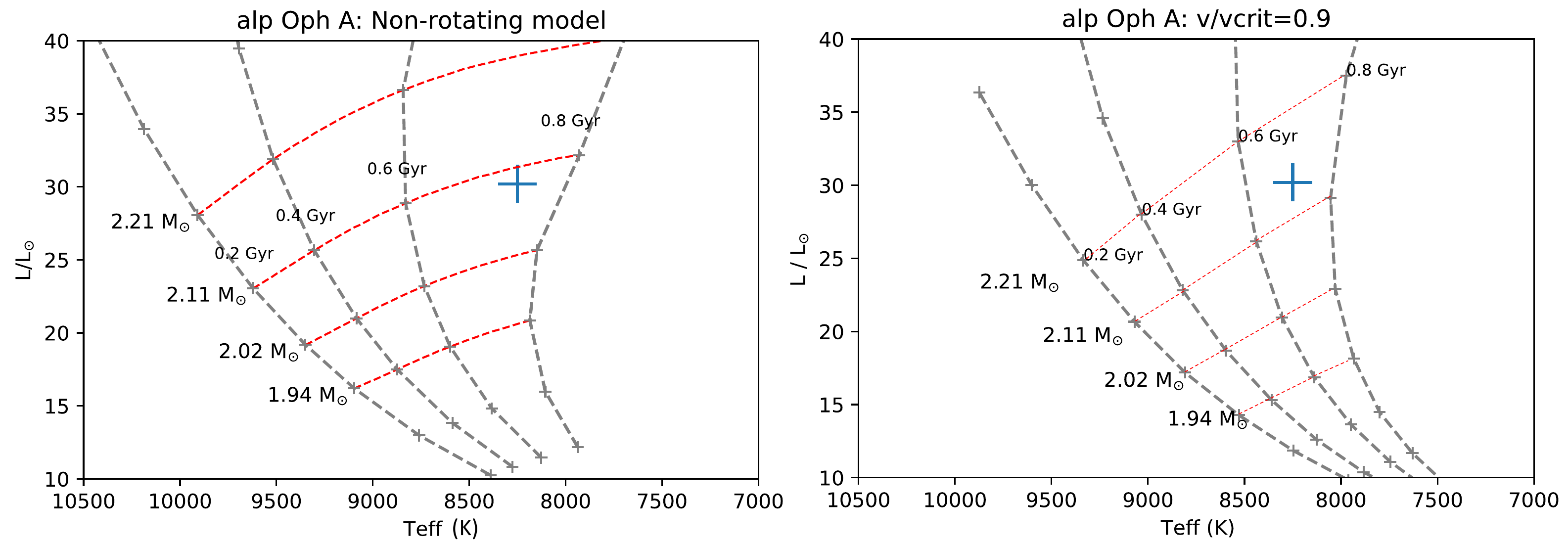}
\caption{We use stellar models with and without rotation in order to plot $\alpha$ Oph A onto an HR diagram. We show MIST tracks without rotation (left) and \citet{ekstrom2012} tracks with $v/v_{\rm{crit}}=0.9$ (right). The stellar track with rotation agrees well with our measured mass, and the isochrone implies an age of 0.7 Gyr. The MIST model has more marginal agreement with a 2.11 M$_{\odot}$ track consistent with $\alpha$ Oph's position, a value just outside our 1-sigma error bars. }
\label{hrdiagram}
\end{figure}

In Figure \ref{hrdiagram} it can be seen that $\alpha$ Oph A falls onto a stellar track which agrees with our measured stellar mass within error bars of $2.20\pm0.06$ M$_{\odot}$ with the rotating-star evolution models. The position on an HR diagram gives an age of about 0.7 Gyr for the system. The fast rotator model is a slightly better fit when compared to a non-rotating evolution model. However, in the non-rotating model a stellar track of 2.11 M$_{\odot}$ still goes through our point, which is $<$2-$\sigma$ away from our best-fit value given our error bars on mass. To better distinguish between these models, we need a mass measurement at the $\sim$1\% level. The mass of the primary comes from the period, semi-major axis, and distance measurement (which includes the semi-amplitudes and inclination). Currently, distance is the quantity which is limiting our mass precision to just under 3\%. We only know distance at the $\sim$1\% level from our RV+visual orbit, and the Hipparcos distance has an error bar twice as high as our new value. If the distance were known perfectly (i.e. a fixed quantity), our error on masses would be only 0.6\% given our orbital element precision. Gaia may eventually improve the distance measurement to this system, though this object does not yet have a distance measurement from Gaia. In fact, $\alpha$ Oph is so bright that the current final parallax precision acheived by the end of the Gaia mission is currently unknown. Precision parallax for bright stars was once thought to be unachievable, though many saturation problems have been solved which gives some hope for these brighter targets \citep{sahlmann2016}. If precision on distance can be improved, we will be able to distinguish between rotating vs non-rotating models on the HR diagram with our precision orbit.

The secondary component of $\alpha$ Oph does not have a literature value for temperature or luminosity. To measure these values, we use our measured J and K flux ratios from Keck AO to compute apparent magnitudes in these bands (presented in Table \ref{physicalelements:alpoph}). For apparent magnitudes of the system, \citet{cohen1999} measure $J=1.728\pm0.005$ and $K=1.639\pm0.005$. These values also match those measured in \citet{alonso1994} within error bars. Though we fit a flux ratio in H-band from MIRC data, we have low confidence in its accuracy for precise photometry. This is due to the fact that the companion is near the edge of the interferometric field-of-view for these epochs, where bandwidth smearing can bias the flux ratio measurement. Hence we choose to only report the J and K band photometry. 

\begin{figure}[H]
\centering
\includegraphics[width=7in]{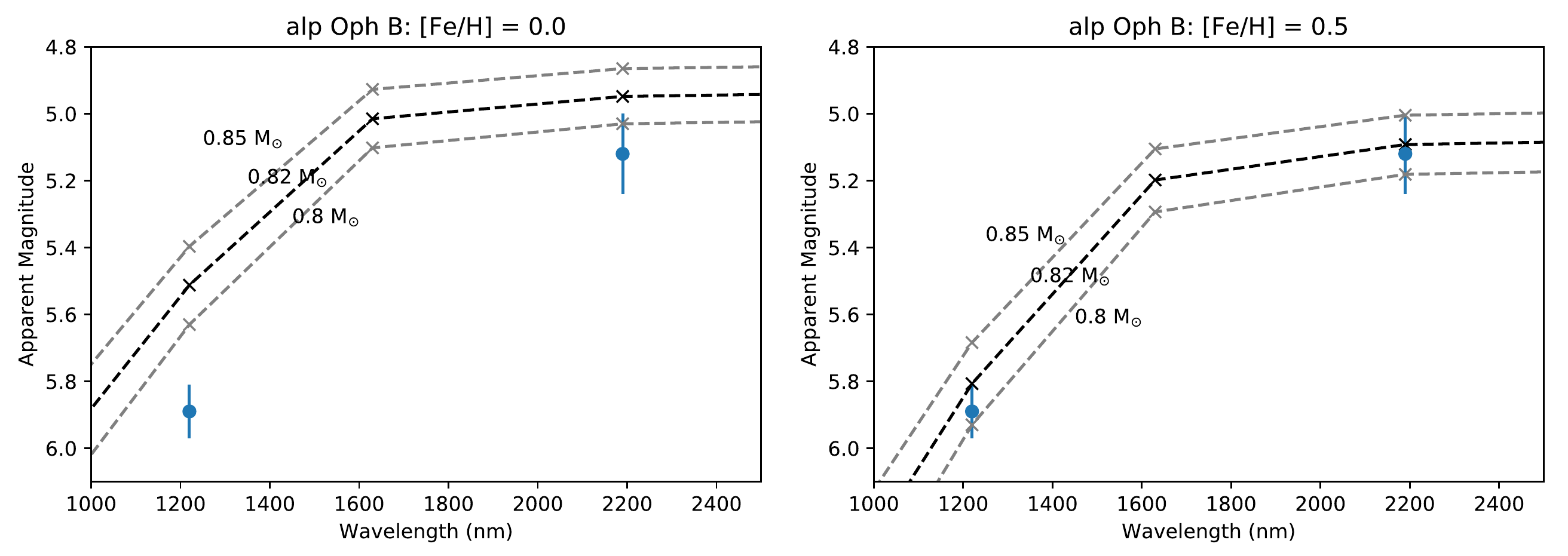}
\caption{We use apparent J and K magnitudes to fit a stellar model for $\alpha$ Oph B. (Left panel) Using MIST models of solar metallicity along with the age and distance measured for the system, we plot the tracks which are consistent with our measured mass and its 1-$\sigma$ uncertainty. This companion is slightly underluminous when compared to stellar models. A larger distance, lower mass of the secondary, or higher metallicity (plotted on right panel) would bring the MIST models into better agreement with our measured mass. }
\label{alpophB_sed}
\end{figure}

We compute synthetic photometry for the secondary using the MIST models at our measured mass and at solar metallicity. Since we have a measurement of distance for the system and age of the primary, we fix these values. Figure \ref{alpophB_sed} shows the results of this fit. We find that our J-band photometry for component B is slightly under-luminous compared to models with our precisely measured mass for $\alpha$ Oph B. Our fit to the measured apparent magnitudes improves if either 1) the distance is higher, 2) the mass of the secondary is lower, or 3) the metallicity is higher. Due to our high precision orbit, options (1) and (2) seem unlikely. Although we note the potential blending issues of the RV data explained in section \ref{sec:observations} which might affect the measured mass ratio. Our mass for the primary is also consistent with that obtained by \citet{hinkley2011} however, though we note that \citet{gatewood2005} obtained a slightly lower mass of $0.7778\pm0.058$ M$_{\odot}$ for the secondary. We show in Figure \ref{alpophB_sed} that a higher metallicity value for $\alpha$ Oph B would make the photometry more consistent with our measured mass. A measurement of metallicity of this component is needed in order to check this consistency.

\section{Establishing $\alpha$ Oph as an Astrometric Reference}
\label{sec:astrometric}

\begin{figure}[H]
\centering
\includegraphics[width=4.5in]{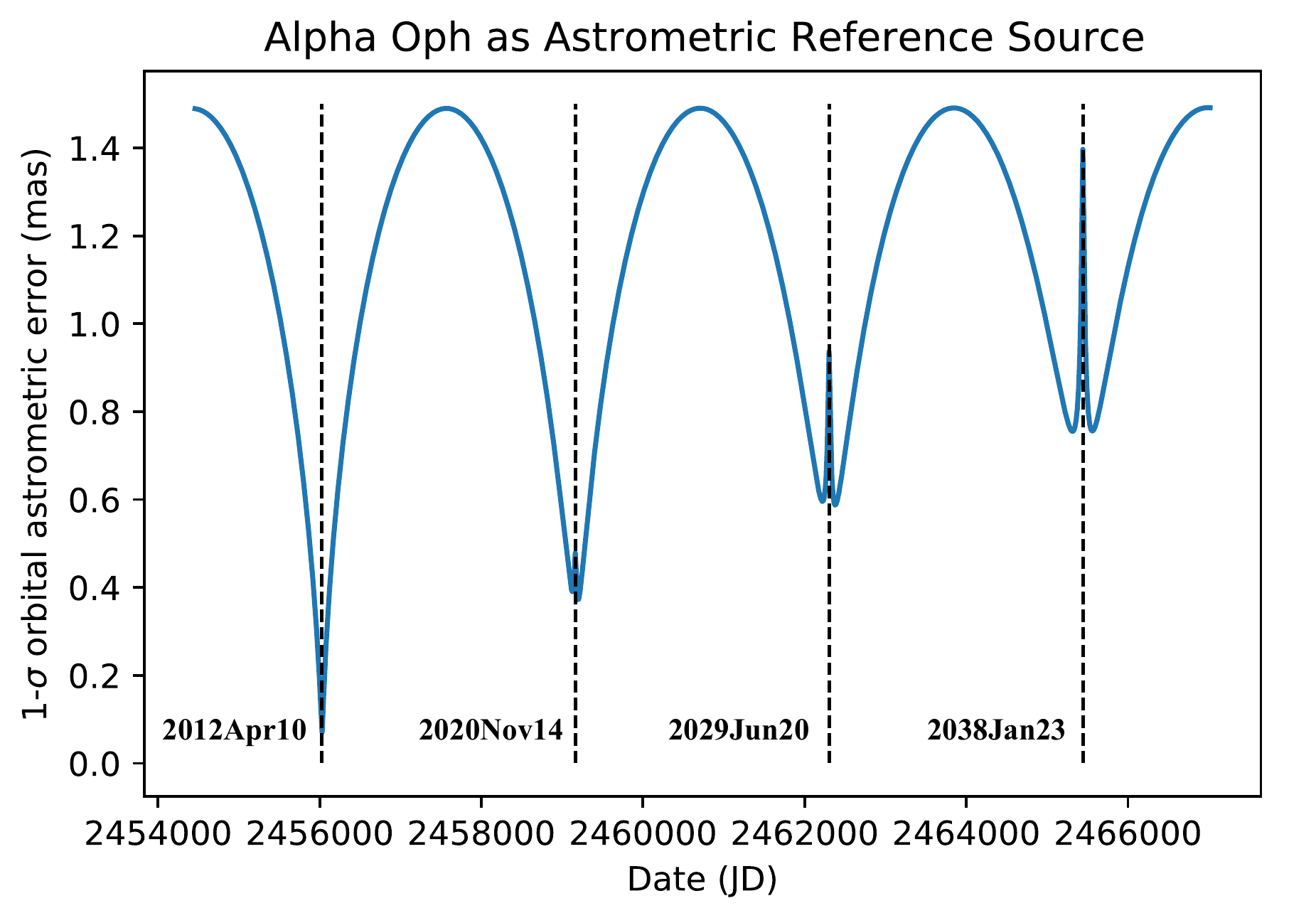}
\caption{We plot the 2D positional error as a function of observing data for our well-constrained orbit of $\alpha$ Oph. The curve shows our 1-$\sigma$ spread about the best solution at a given date using the posterior chains from the MCMC routine. Astrometric prediction across the next decade are accurate to $<$1.5 mas, with data near periastron approaching the 100s of $\mu$-as level. Additional high precision data outside of periastron passage will improve this source as a reference over the coming decades. }
\label{alpoph_reference}
\end{figure}

Our new data of the $\alpha$ Oph binary system near periastron passage increases the precision of the system's orbital elements, making this bright source a potential high precision reference source for astrometric programs. High precision binary orbits are useful as calibration sources for instruments to measure wavelength or astrometry calibration. Since wider binaries often have long orbital periods, precision orbits at the level needed by such instruments are often sparse. In Figure \ref{alpoph_reference} we demonstrate our current astrometric precision for this system given our current orbital parameters and their associated uncertainties. We sample orbits from our MCMC posterior distribution and report the 1-$\sigma$ spread about the best-fit orbit at a given time over the next decade. As can be seen in the figure, our orbital element precision was at the $<$200 $\mu$-as level during the 2012 periastron passage, at $\sim$400 $\mu$-as during the 2020 passage, and at a 600-900 $\mu$-as level for the 2029 periastron passage. Our orbital elements lead to astrometry predictions at a $<$1.6 mas spread across the 8.6-year orbit, with the more precise predictions near periastron passage. This orbital precision can be improved by taking more high precision data outside of periastron passages, which could make this bright binary source an even better astrometric calibration source in the future.

\section{Conclusions}
\label{sec:conclusions}
We obtained new interferometric and adaptive optic imaging data on the nearby, bright A5IV binary star $\alpha$ Oph. We also report a new double-lined RV orbit which covers two periastron passages of this eccentric system. Since this binary system is extremely eccentric (e = 0.94), data near periastron passage are crucial for improving the orbital elements and mass measurements of this $\sim$8.6 year period system. Measuring the primary mass at a high precision is especially desired for $\alpha$ Oph, as this star is also a rapid rotator which has been imaged and modeled with high angular resolution interferometric measurements. Visual and RV orbits of binary stars are the only way to directly measure stellar masses to high precision, and we use our well-covered orbit of $\alpha$ Oph to validate mass estimates which come from the modeling of rapidly rotating stars. This system is highly unique in being an imaged rapid rotator and having a well characterized binary orbit. Other imaged rotators such as $\alpha$ Cep and $\beta$ Cas do not have known companions, while the companion for the rapid rotator Regulus is complicated by mass transfer and extremely hard to detect as a visual binary. $\alpha$ Oph provides a benchmark test of rotating evolution models that is not easy to achieve with other known systems. Our updated orbit allows for a strict comparison with rapid rotator models. 

\citet{hinkley2011} previously used adaptive optics imaging data to present an orbit and masses for this system, and that paper called for additional data near periastron passage to better constrain the masses. Our previously unpublished data near periastron passage, as well as a full RV orbit, now allows for high precision comparison of masses with rotator models. We directly measure a primary mass of $2.20\pm0.06$ M$_{\odot}$, in agreement with the model predictions using rapid rotator imaging results of \citet{zhao2009}. We compare the primary star with stellar evolution models on an HR diagram, and find that it fits well on a track with our measured stellar mass, solar metallicity, and an age of $\sim$0.7 Gyr. We find that stellar evolution models including rotation give a slightly more consistent fit to our measured mass, though even with our 2.7\% mass error this is a $<$2-$\sigma$ result. To better distinguish between evolution models, we need $\sim$1\% mass errors. This can be accomplished by improving the error on distance to the system, which Gaia might be able to do in future data releases. The secondary component of the binary system is slightly underluminous for our measured mass of $0.824 \pm0.023$ M$_{\odot}$. This discrepancy might be explained if the metallicity is higher than solar. The mutual inclination of the rotation axis of the star and the orbital plane of the binary is non co-planar at either $42.7 \pm 0.6^{\circ}$ or $133.0 \pm 0.6^{\circ}$, depending on the rotation orientation of the primary. The mutual inclination and high eccentricity of the binary system may hint at early interactions in the star formation stages. 

We demonstrate that current orbital precision makes $\alpha$ Oph a potentially useful astrometric calibration source for other instruments. It is bright, easily observable, and the binary can be resolved by single-aperture telescopes. Additional high precision astrometry data away from periastron could improve the orbital elements further, to approach $<$100 $\mu$-as predictive power over the next orbit. 

\acknowledgments
T.G. and J.D.M. acknowledge support from NASA-NNX16AD43G, and from NSF-AST2009489. T.G. acknowledges support from Michigan Space Grant Consortium, NASA grant NNX15AJ20H. S.K. acknowledges support from an ERC Starting Grant (grant agreement no. 639889). Astronomy at
Tennessee State University is supported by the state of
Tennessee through its Centers of Excellence program.
This work is based upon observations obtained with the Georgia State University Center for High Angular Resolution Astronomy Array at Mount Wilson Observatory.  The CHARA Array is supported by the National Science Foundation under Grant No. AST-1636624 and AST-1715788.  Institutional support has been provided from the GSU College of Arts and Sciences and the GSU Office of the Vice President for Research and Economic Development. 
Some of the data presented herein were obtained at the W. M. Keck Observatory, which is operated as a scientific partnership among the California Institute of Technology, the University of California and the National Aeronautics and Space Administration. The Observatory was made possible by the generous financial support of the W. M. Keck Foundation. This research has made use of the Keck Observatory Archive (KOA), which is operated by the W. M. Keck Observatory and the NASA Exoplanet Science Institute (NExScI), under contract with the National Aeronautics and Space Administration.
The authors wish to recognize and acknowledge the very significant cultural role and reverence that the summit of Maunakea has always had within the indigenous Hawaiian community.  We are most fortunate to have the opportunity to conduct observations from this mountain.
This research has made use of the Jean-Marie Mariotti Center SearchCal service\footnote{available at \url{http://www.jmmc.fr/searchcal_page.htm}}.

% * <monnier@umich.edu> 2017-08-07T14:50:24.545Z:
% 
% include chara grants and acknowledgements from others. if some authors drop off the list, also thank them here
% 
% ^.


%% for the paper.  Each keyword is check against the master list during
%% copy editing.  Individual instruments can be provided in parentheses,
%% after the keyword, but they are not verified.

\vspace{5mm}
\facilities{CHARA, Fairborn Observatory, Keck:II (NIRC2)}
\software{lmfit, astropy, emcee}



\bibliographystyle{aasjournal}
\bibliography{references}

\listofchanges

\appendix
\section{Radial Velocity Data}
\label{appendix:rv}
%RV table
\begin{longtable}{lcc}
\caption{$\alpha$ Oph Radial Velocities} \\
%\label{table:rv-alpoph}\\
\hline
HJD\tablenotemark{a} & $RV_A$ (km~s$^{-1}$)\tablenotemark{b} & $RV_B$ (km~s$^{-1}$)\tablenotemark{c} \\
\hline
\endhead
2455847.6533 & 7.8 & -- \\
2455927.0572 & 10.2 & -- \\
2455930.0570 & 10.0 & -- \\
2455935.0558 & 11.5 & -- \\
2455947.0524 & 12.1 & -- \\
2455957.0473 & 11.9 & -- \\
2455967.0446 & 10.9 & -- \\
2455976.0409 & 10.8 & -- \\
2455991.0334 & 12.4 & -- \\
2456001.0269 & 14.8 & -- \\
2456017.9313 & 22.2 & -21 \\
2456019.9748 & 23.0 & -23 \\
2456022.8985 & 25.0 & -- \\
2456025.9039 & 32.1 & -44 \\
2456028.0124 & 33.3 & -- \\
2456029.8003 & 30.9 & -51 \\
2456029.9008 & 31.8 & -- \\
2456030.8395 & 31.4 & -50 \\
2456030.9554 & 30.8 & -49 \\
2456032.9811 & 28.8 & -44 \\
2456033.8286 & 31.3 & -39 \\
2456035.0076 & 31.1 & -- \\
2456037.0024 & 27.6 & -32 \\
2456038.0022 & 26.5 & -30 \\
2456045.0015 & 21.2 & -- \\
2456046.0008 & 19.4 & -- \\
2456046.9992 & 20.0 & -- \\
2456048.9992 & 20.3 & -- \\
2456050.9982 & 18.1 & -- \\
2456051.9965 & 18.8 & -8 \\
2456052.9940 & 19.4 & -- \\
2456055.9947 & 17.6 & -- \\
2456060.9922 & 17.1 & -- \\
2456061.8736 & 19.1 & -3 \\
2456063.8189 & 15.8 & -3 \\
2456065.8098 & 16.6 & -2 \\
2456067.7947 & 16.0 & -1 \\
2456074.7881 & 15.0 & 0 \\
2456077.9448 & 15.0 & 0 \\
2456078.9837 & 15.1 & -- \\
2456081.9150 & 14.2 & 1 \\
2456087.7671 & 15.0 & -1 \\
2456092.8878 & 13.7 & 0 \\
2456134.8539 & 10.8 & -- \\
2456135.6749 & 11.6 & -- \\
2456170.7241 & 10.6 & -- \\
2456172.7161 & 10.9 & 10 \\
2456185.6536 & 9.4 & 10 \\
2456188.6476 & 10.7 & 6 \\
2456189.7018 & 11.9 & 9 \\
2456191.6455 & 9.8 & 8 \\
2456192.6322 & 11.5 & -- \\
2456193.6779 & 10.9 & -- \\
2456194.6758 & 11.1 & 10 \\
2456195.6820 & 10.4 & 9 \\
2456196.6412 & 11.6 & -- \\
2456197.7243 & 10.4 & 7 \\
2456226.5981 & 11.6 & 12 \\
2456227.5979 & 10.5 & 13 \\
2456228.5831 & 9.9 & 12 \\
2456233.5810 & 10.2 & 11 \\
2456234.5802 & 9.5 & 10 \\
2456235.5796 & 11.3 & -- \\
2456236.5789 & 10.3 & 11 \\
2456387.9042 & 10.7 & 15 \\
2456388.7848 & 9.4 & 16 \\
2456389.7884 & 9.6 & 15 \\
2456392.7780 & 9.1 & 14 \\
2456464.7862 & 9.1 & 14 \\
2456465.7862 & 9.3 & 16 \\
2456466.7862 & 9.9 & 16 \\
2456559.6290 & 5.2 & 12 \\
2456560.6282 & 4.9 & 13 \\
2456568.6220 & 6.1 & 12 \\
2456569.6218 & 5.9 & 12 \\
2456750.8676 & 10.2 & 15 \\
2456751.7894 & 8.5 & 18 \\
2456752.8030 & 8.9 & 17 \\
2456753.8034 & 10.2 & 18 \\
2456754.7838 & 8.9 & 19 \\
2456783.8659 & 10.7 & 16 \\
2456813.9677 & 12.8 & 16 \\
2457062.9706 & 7.2 & 13 \\
2457291.6473 & 6.4 & 20 \\
2457328.5954 & 5.7 & 19 \\
2457329.6022 & 5.2 & 17 \\
2457331.5911 & 5.0 & 17 \\
2457472.9618 & 8.2 & 15 \\
2457479.8958 & 8.7 & 15 \\
2457481.9314 & 8.2 & 16 \\
2457487.8848 & 9.0 & 16 \\
2457493.8473 & 6.8 & 13 \\
2457509.7589 & 7.4 & 16 \\
2457528.9745 & 8.1 & 15 \\
2457544.9048 & 9.0 & 15 \\
2457579.9297 & 6.9 & -- \\
2457666.6528 & 5.8 & 18 \\
2457916.8663 & 8.8 & 15 \\
2458003.7437 & 5.4 & 19 \\
2458043.6544 & 6.6 & -- \\
2458263.8664 & 9.7 & 15 \\
2458269.6760 & 8.7 & 15 \\
2458270.8365 & 10.0 & 18 \\
2458274.8666 & 9.3 & 17 \\
2458275.8466 & 8.4 & 18 \\
2458277.8466 & 9.8 & 19 \\
2458287.8366 & 8.4 & 16 \\
2458288.8366 & 8.2 & 16 \\
2458628.6836 & 5.7 & 18 \\
2458629.6841 & 5.8 & 17 \\
2458630.6872 & 5.9 & 17 \\
2458652.7964 & 5.8 & 18 \\
2458653.7964 & 6.4 & 19 \\
2458654.7964 & 6.9 & 18 \\
2458968.7391 & 10.0 & 12 \\
2458972.9149 & 9.8 & 15 \\
2459020.6609 & 10.5 & 16 \\
2459021.6614 & 10.7 & 18 \\
2459022.6610 & 9.7 & 18 \\
2459097.6584 & 11.3 & 11 \\
2459098.6461 & 9.9 & 8 \\
2459100.6702 & 9.9 & 11 \\
2459103.6290 & 9.1 & 9 \\
2459108.6256 & 10.0 & 10 \\
2459109.6239 & 10.6 & 11 \\
2459110.6197 & 10.5 & -- \\
2459131.6945 & 11.3 & -- \\
2459132.5989 & 14.1 & 8 \\
2459133.5976 & 13.9 & -- \\
2459134.5970 & 13.9 & 2 \\
2459135.5958 & 12.4 & 1 \\
2459136.6397 & 13.3 & 1 \\
2459146.6325 & 15.2 & -2 \\
2459170.5849 & 33.5 & -52 \\
2459171.5639 & 32.2 & -- \\
2459172.5635 & 30.3 & -42 \\
2459173.5632 & 29.6 & -41 \\
2459174.5625 & 27.7 & -39 \\
2459175.5622 & 26.7 & -35 \\
2459178.5577 & 24.9 & -29 \\
2459179.5566 & 26.0 & -29 \\
2459180.5549 & 24.1 & -25 \\
2459181.5536 & 23.9 & -23 \\
2459182.5519 & 25.1 & -- \\
2459183.5501 & 23.5 & -- \\
\hline
\end{longtable}
\tablenotetext{a}{HJD = Heliocentric Julian Date}
\tablenotetext{b}{Errors on primary are 1.4 km~s$^{-1}$.}
\tablenotetext{c}{Errors on secondary are 2 km~s$^{-1}$.}

\begin{figure}[H]
\centering
\includegraphics[width=4.5in]{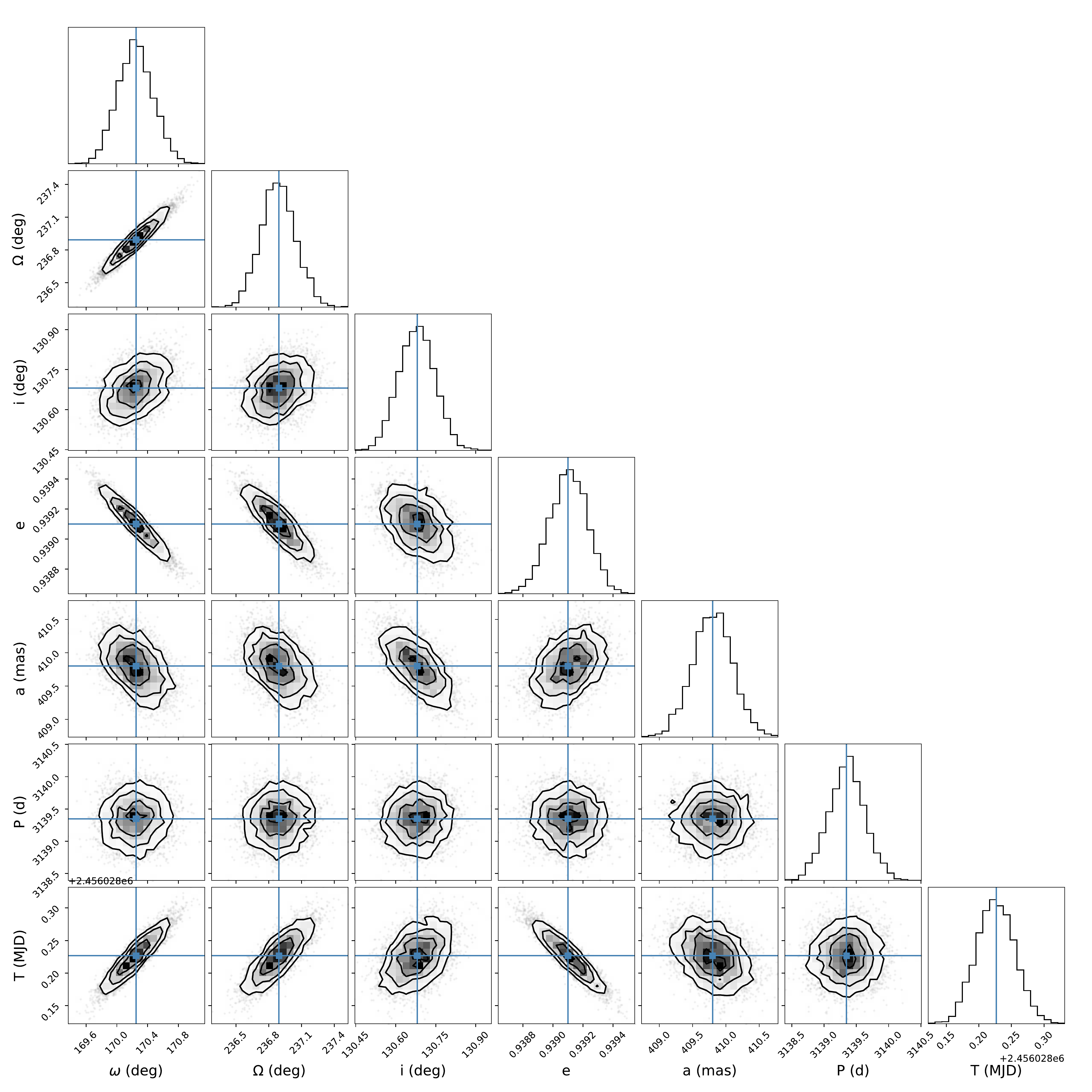}
\caption{We show the corner plot of posterior distributions from our MCMC routine to estimate errors on the orbital parameters. Since the RV parameters do not show significant correlations, we include only the visual orbital elements for clarity. There are correlations of the time of periastron passage with the elements of eccentricity, $\omega$, and $\Omega$ which is normal for eccentric orbits. }
\label{alpoph_mcmc}
\end{figure}

\end{document}